\documentclass[pdftex,final,twocolumn,epjc3]{svjour3}

\RequirePackage[utf8]{inputenc}
\RequirePackage{graphicx}
\RequirePackage{mathptmx}
\RequirePackage{flushend}
\RequirePackage[colorlinks,citecolor=blue,urlcolor=blue,linkcolor=blue]{hyperref}
\RequirePackage{xspace}
\RequirePackage{caption}
\RequirePackage{subcaption}
\RequirePackage{amsmath}
\RequirePackage[protrusion=true,expansion]{microtype}
\RequirePackage{makecell}

\newcommand{\pT}{\ensuremath{p_{\textrm{\scriptsize{T}}}}\xspace}

\usepackage[url=false,doi=true,backend=biber,bibencoding=utf8,sorting=none,style=numeric-comp]{biblatex}
\journalname{}
\begin{document}
\title{Development of a resource-efficient FPGA-based neural network regression model for the ATLAS muon trigger upgrades}

\author{Rustem~Ospanov\thanksref{e1,a1} \and
    Changqing~Feng\thanksref{e2,a1}
    \and
    Wenhao~Dong\thanksref{a1}
    \and
    Wenhao~Feng\thanksref{a1}
    \and 
    Kan~Zhang\thanksref{a1}
    \and 
    Shining~Yang\thanksref{a1}
}

\thankstext{e1}{e-mail: rustem@cern.ch}
\thankstext{e2}{e-mail: fengcq@ustc.edu.cn}

\institute{Department of Modern Physics and State Key Laboratory of Particle Detection, University of Science and Technology of China, Hefei\label{a1}}

\date{\today}
\maketitle
\begin{abstract}
   This paper reports on the development of a resource-efficient FPGA-based neural network regression model for potential applications in the future hardware muon trigger system of the ATLAS experiment at the Large Hadron Collider (LHC). Effective real-time selection of muon candidates is the cornerstone of the ATLAS physics programme. With the planned ATLAS upgrades for the High Luminosity LHC, an entirely new FPGA-based hardware muon trigger system will be installed that will process full muon detector data within a 10~${\mu}s$ latency window. The large FPGA devices planned for this upgrade should have sufficient spare resources to allow deployment of machine learning methods for improving identification of muon candidates and searching for new exotic particles. Our neural network regression model promises to improve rejection of the dominant source of background trigger events in the central detector region, which are due to muon candidates with low transverse momenta. This model was implemented in FPGA using 157 digital signal processors and about 5,000 lookup tables. The simulated network latency and deadtime are 122 and 25~ns, respectively, when implemented in the FPGA device using a 320~MHz clock frequency. Two other FPGA implementations were also developed to study the impact of design choices on resource utilisation and latency. The performance parameters of our FPGA implementation are well within the requirements of the future muon trigger system, therefore opening a possibility for deploying machine learning methods for future data taking by the ATLAS experiment.
\end{abstract}

\section{Introduction} 

Machine learning methods have become standard practice in experimental particle physics, see Refs.~\cite{nature,Carleo:2019ptp} for recent reviews. They are frequently used to detect rare processes in offline data analysis, where they are typically executed on servers utilising Intel and AMD processor architectures. Recently, several different machine learning methods have been implemented using field-programmable gate array (FPGA) devices for real-time applications in high energy physics~\cite{Duarte:2018ite,Nottbeck:2019rqu,Coelho:2020zfu,Hong:2021snb,Govorkova:2021utb,epjcroma,CMS:2018wav,Sun:2022bxx}. The current state of applications and techniques for fast machine learning is summarised in the recently released community report~\cite{Deiana:2021niw}. 

FPGA-based machine learning algorithms are deployed in situations where software solutions cannot provide sufficiently high bandwidth within a required low latency. For example, experimental trigger systems at the Large Hadron Collider (LHC) are a perfect use case for deploying FPGA-based neural networks. These systems use sophisticated data filtering algorithms (triggers) to select in real-time interesting collision events. First-level hardware components of these systems process a large amount of data at a 40~MHz rate of LHC collisions with microsecond-level latency, thus necessitating hardware-based machine learning solutions. For instance, FPGA-based algorithms have been proposed for searching for new physics phenomena using calorimeter data~\cite{Alimena:2020web,Linthorne:2021oiz} and kinematic properties of events~\cite{Cerri:2018anq}. FPGAs are also used to accelerate resource-intensive software algorithms for processing data by large server farms~\cite{Duarte:2019fta}. Such accelerators are being studied for potential deployment in high-level software components of the trigger systems at the LHC.

The present work is motivated by future availability of large FPGA devices that will be installed during the planned upgrades of the ATLAS muon trigger system~\cite{CERN-LHCC-2017-020,CERN-LHCC-2017-017}. These FPGA devices will allow deployment of more sophisticated real-time selection algorithms than what is currently possible. Our goal is the development of generic FPGA-based neural network algorithms that can be used to extend physics capabilities of the future ATLAS trigger system. An advantage of using neural networks is that they are capable of accurately approximating multivariate functions~\cite{Barron93,HornikSW89}. Therefore, the same neural network circuit design can be used for different applications, with differences configured via different network parameter sets. Deploying such FPGA-based machine learning methods for real-time applications requires careful optimisations in order to reach necessary timing and resource utilisation targets. Performing these optimisations in the context of the ATLAS muon trigger upgrades is one of the subjects of the present work.

In this paper, a resource-efficient FPGA-based neural network regression model is developed that meets FPGA resource and latency requirements of the future ATLAS muon trigger system~\cite{CERN-LHCC-2017-020}. These FPGA-based networks will be used to improve performance of standard muon trigger algorithms and to search for new exotic particles. For example, a new trigger can be developed to search for slow-moving heavy charged particles using the resistive plate chamber (RPC) detector~\cite{RPCperf} for time-of-flight measurements~\cite{Aaboud:2019trc}. Deploying such triggers for analysing collision data will improve sensitivity of the LHC experiments for detecting new long-lived particles that are predicted by well-motivated extensions of the standard model (e.g., Refs.~\cite{Alimena:2019zri,Curtin:2018mvb}). The current RPC trigger system~\cite{Anulli:2009zz,CERN-LHCC-2017-017} does not allow such triggers because it uses application-specific integrated circuits that perform standard muon logic. The current system will be replaced with an FPGA-based system after 2026. Our goal is to deploy resource-efficient neural networks using spare FPGA resources that should be available after implementation of baseline muon trigger algorithms~\cite{CERN-LHCC-2017-020} that do not use machine learning methods.

This paper reports on the development of an FPGA-based neural network regression model that aims to improve performance of the RPC trigger system by measuring more precisely muon transverse momentum ($\pT$) and charge. This model shows promising potential for reducing rates of hardware-based muon trigger algorithms of the future ATLAS trigger system. It was implemented in FPGA code using the hardware description language (HDL). The HDL is used in order to minimise usage of FPGA resources and to achieve better timing performance.  Two other FPGA implementations were also developed to study the impact of design choices on resource utilisation and latency. As detailed later, the resource usage, latency and deadtime of our model are well within the requirements of the future ATLAS hardware trigger system. In this context, the deadtime denotes a time period during which an FPGA circuit is busy with calculations and therefore is not available for processing new data. This result creates an opportunity for deploying novel hardware-based trigger algorithms that use machine learning methods with a minimal impact on the new ATLAS trigger system, which is still being designed. 

This paper is organised as follows. Section~\ref{sec:atlas} describes the ATLAS detector and RPC muon trigger system. This section describes key detector features that are included in the simple RPC detector simulation model described in Section~\ref{sec:design}. This section also describes the design of the neural network regression model for measuring muon candidate $\pT$ and charge. Section~\ref{sec:result} details performance of this model and compares its efficiency for detecting muon candidates with that of the present ATLAS RPC muon trigger. Section~\ref{sec:fpga} details design and performance of the FPGA implementation of this model. Section~\ref{sec:sum} summarises our results and ideas for the future work.

\section{ATLAS experiment and muon trigger system} 
\label{sec:atlas}

The ATLAS experiment at the LHC is a general purpose detector observing high energy collisions of protons and heavy ions. The detector is designed for efficient detection of leptons, hadronic jets and missing transverse energy. The ATLAS physics programme includes measurements of the Higgs boson properties, discovered simultaneously with CMS in 2012~\cite{Aad20121,Chatrchyan201230}, measurements of standard model properties, and diverse searches for new physics phenomena. Efficient selection of muon candidates is the crucial requirement of the ATLAS physics programme. 

The ATLAS detector~\cite{atlas-detector} consists of several sub-detectors with cylindrical geometry. The LHC beamline serves as the detector $z$-axis. The detector consists of one central barrel section and two endcap sections. The inner tracking detectors are immersed in 2~T magnetic field allowing precise measurements of the charged particle momenta. The electromagnetic and hadronic calorimeters are located outside the tracking detectors. The muon spectrometer is located outside the calorimeters and immersed in approximately 0.5~T magnetic field generated by three air-core toroidal magnets.

Interesting collision events are selected in real-time by the two-level trigger system. The first level trigger system (L1) uses dedicated hardware algorithms to analyse data from fast muon detectors and partial calorimeter data, operating at the 40~MHz LHC collision rate. The L1 system accepts events at approximately 100~kHz rate for further analysis by a high-level software-based trigger system which uses software algorithms to select events for offline analysis at approximately 1~kHz rate. 

The current L1 system uses the RPC detector to select muon candidates in the central (barrel) region of the detector. The RPCs are fast gaseous detectors~\cite{Santonico:1981sc,Santonico:1988qi,RPCperf} with space and time resolution of about 1~cm and 1~ns, respectively. The present ATLAS RPCs are arranged into three concentric double layers (doublets) that are located at radii of approximately 6.8~m, 7.5~m and 9.8~m, referred to as RPC1, RPC2 and RPC3, respectively. The RPC-based muon trigger algorithms were implemented using application-specific integrated circuits that were developed specifically for the RPC trigger system~\cite{Anulli:2009zz}.

The L1 muon trigger identifies muon candidates and measures their $\pT$ using six thresholds~\cite{Aad:2020uyd}. The primary single muon trigger (MU20) corresponds to the $\pT$ threshold of 20~GeV. This trigger selects events at about 15~kHz rate at the highest instantaneous LHC luminosity achieved in 2018. The majority of muons produced in LHC collisions are due to decays of heavy flavour hadrons and decays of $W$ and $Z$ bosons~\cite{Aad:2011rr}. The majority of the selected L1 trigger muon candidates are low-$\pT$ muons with mismeasured $\pT$ values~\cite{RPCperf}. 

\begin{figure}[ht]
    \centering
    \includegraphics[width=0.49\textwidth]{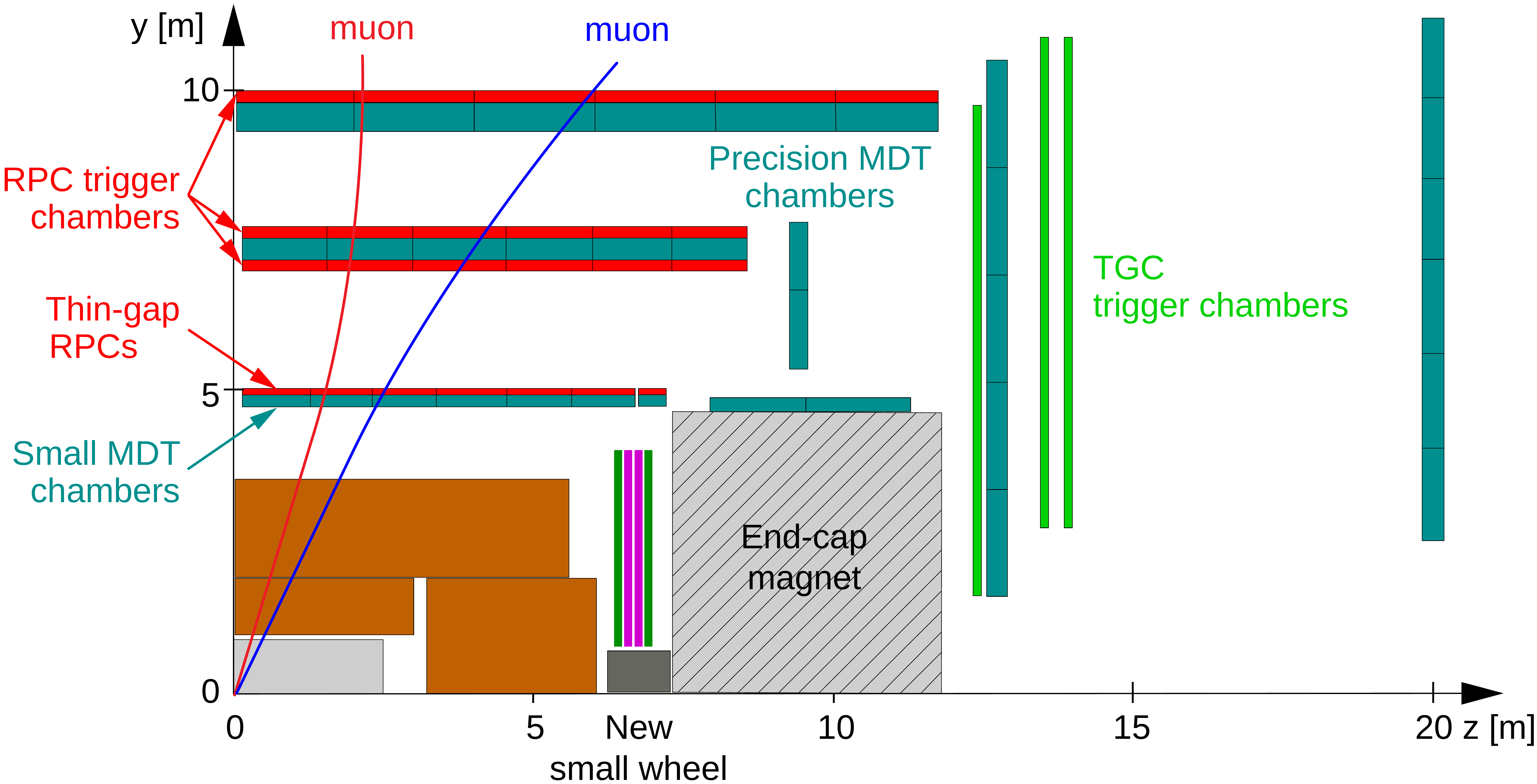}
    \caption{One quadrant of the ATLAS muon spectrometer after the upgrades which will add new thin-gap RPCs and MDTs in the inner barrel region.  The $z$-axis points along the beam direction and the collision point is at the origin.}
    \label{fig:det}
\end{figure}

The ATLAS experiment will undergo extensive upgrades over the next several years to prepare for higher collision rates of the High Luminosity LHC, and correspondingly higher backgrounds. In order to maintain efficient trigger selections, an entirely new hardware trigger system will be installed that will select interesting LHC collisions at about 1~MHz rate~\cite{CERN-LHCC-2017-020}. New muon trigger algorithms will be executed using powerful FPGA devices which will allow more complex, re-programmable trigger logic~\cite{CERN-LHCC-2017-020}. As a part of these upgrades, new thin-gap RPCs and small diameter muon drift tubes (MDTs) will be also installed in the inner barrel region of the muon spectrometer~\cite{CERN-LHCC-2017-017}. Figure~\ref{fig:det} shows the ATLAS muon spectrometer after the upgrades.

\section{RPC simulation model and neural network design}
\label{sec:design}

A simple simulation model of the current RPC detector has been developed for designing and testing a neural network regression model~\footnote{\href{https://github.com/rustemos/MuonTriggerPhase2RPC}{https://github.com/rustemos/MuonTriggerPhase2RPC}} for measuring muon $\pT$. The geometry of the present detector is used for this study to allow performance comparisons with the current RPC muon trigger~\cite{RPCperf}. This model includes three cylindrical doublet RPC layers, with each doublet layer made of two parallel detector surfaces separated by 2~cm. Active detector elements are simulated by parallel strips that are 3~cm wide. The simulation model includes only so-called $\eta$ strips~\cite{RPCperf} that measure muon deflections in the bending $(r,z)$ plane.

Approximately 100,000 positively and 100,000 negatively charged muons are simulated, originating at the detector centre. Muons are simulated with a uniform \pT distribution in the range of 3~GeV to 30~GeV, and with a uniform distribution of muon angles in the range of 40 to 85 degrees with respect to the $z$-axis. Muons with $\pT<3$~GeV are typically stopped by the calorimeters and therefore they are not included in the training. Muons with $\pT>30$~GeV cannot be resolved by the RPC system as they look like straight tracks, therefore including them does not help to improve the network performance. Muons are propagated through the uniform toroidal magnetic field of 0.5~T in the detector region with a radius greater than 6~m. No material scattering effects are included in the simulation. The effect of the material scattering on muon trajectories is expected to be smaller than the RPC strip width and therefore should not affect our results. Verifying this assumption with more precise simulation will be one of the subjects of follow up work.

\begin{figure}[ht]
    \centering
    \includegraphics[width=0.49\textwidth]{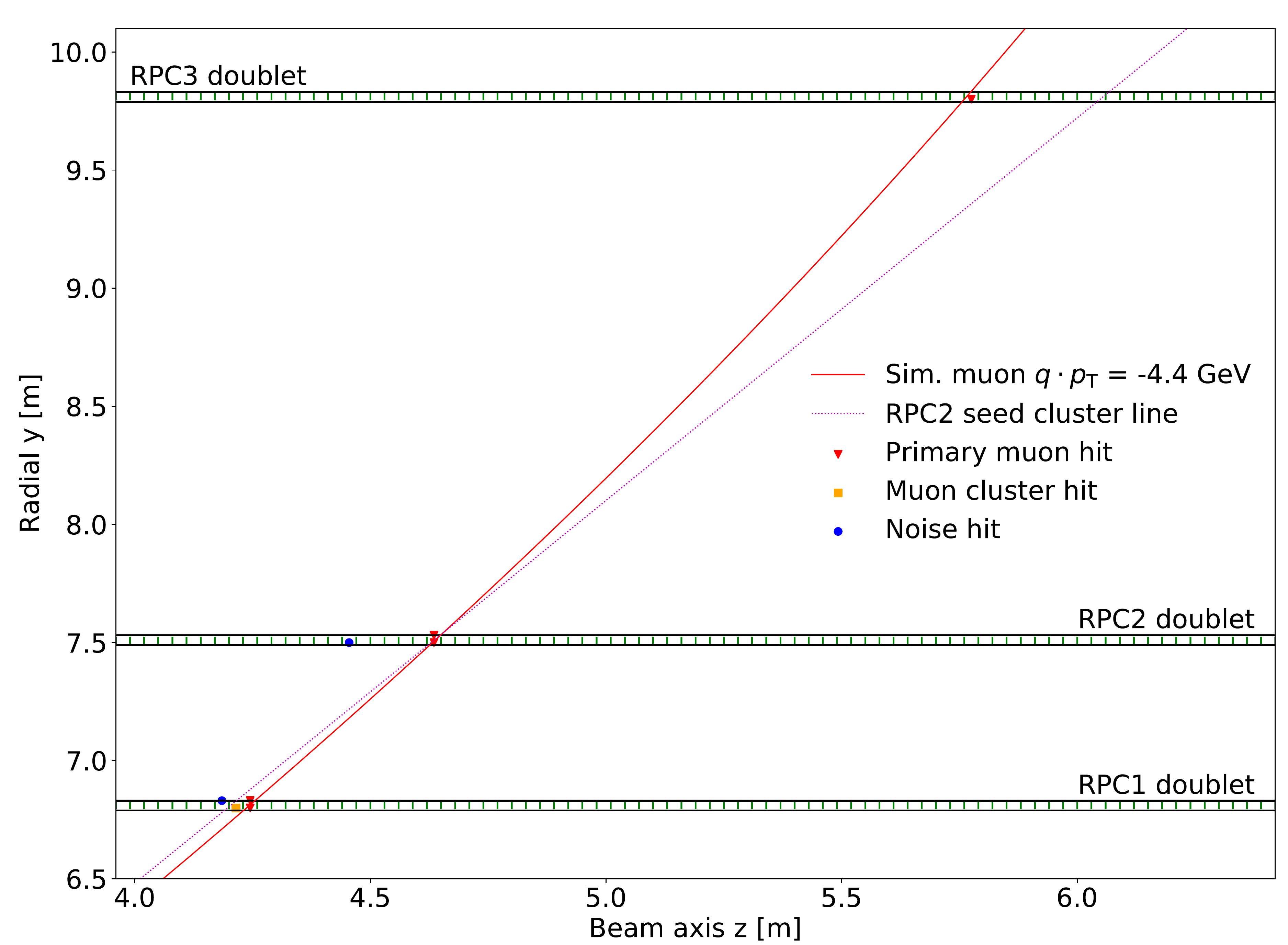}
    \caption{Illustration of one simulated muon candidate traversing the RPC detector. Six horizontal solid black lines represent six layers of the current RPC detector. Small vertical lines along these lines represent strip boundaries. Solid red line represents the muon track. Dotted magenta line passing through the reconstructed cluster in the RPC2 layer is referred to as the seed line in the text. Coloured markers represent simulated RPC hits.}
     \label{fig:muplus}
\end{figure}

The probability for a muon passing through a given strip to produce a hit in that strip is 95\%. In addition, each muon has 25\% average probability to produce one additional hit in the closest strip and 5\% probability to produce two hits in two nearby strips. These additional hits are referred to as cluster hits. The probability to produce an extra (cluster) hit is 0\% at the strip's centre and increases linearly to 50\% at the strip's edge. Finally, the probability to produce a noise (background) hit in each strip is 0.1\%. These noise hits correspond to ionisation events in the real detector due to background particles, such as low momentum photons and neutrons. The above values for cluster and noise hit probability were chosen to approximate the actual detector response~\cite{RPCperf}. Figure~\ref{fig:muplus} shows an event display of one simulated muon traversing the RPC detector model.

Each simulated event is processed to reconstruct clusters using adjacent hits in each doublet layer. A cluster contains all contiguous hits in nearby strips in the two layers belonging to one doublet layer. A candidate muon is required to contain one cluster in each of three RPC layers. The reconstruction algorithm iterates over all clusters in the RPC2 layer, with each cluster serving as a seed for building a muon candidate. The two selected clusters in the RPC1 and RPC3 layers are clusters that are closest to the straight line starting at the origin and passing through the centre of the RPC2 cluster (referred to as the seed line). The cluster centre is the mean position of all cluster hits. The centres of the selected clusters are required to be within a window of 0.15~m and 0.6~m with respect to the seed line for the RPC1 and RPC3 clusters, respectively. These window sizes were chosen to collect the majority of muons with $\pT>20$~GeV that curve in the magnetic field while still remaining within the window. Each event typically contains one muon candidate. For a small fraction of events, two muon candidates are reconstructed when a noise cluster in one layer matches muon induced clusters in other two layers.

\begin{figure}[ht]
    \centering
    \includegraphics[width=0.49\textwidth]{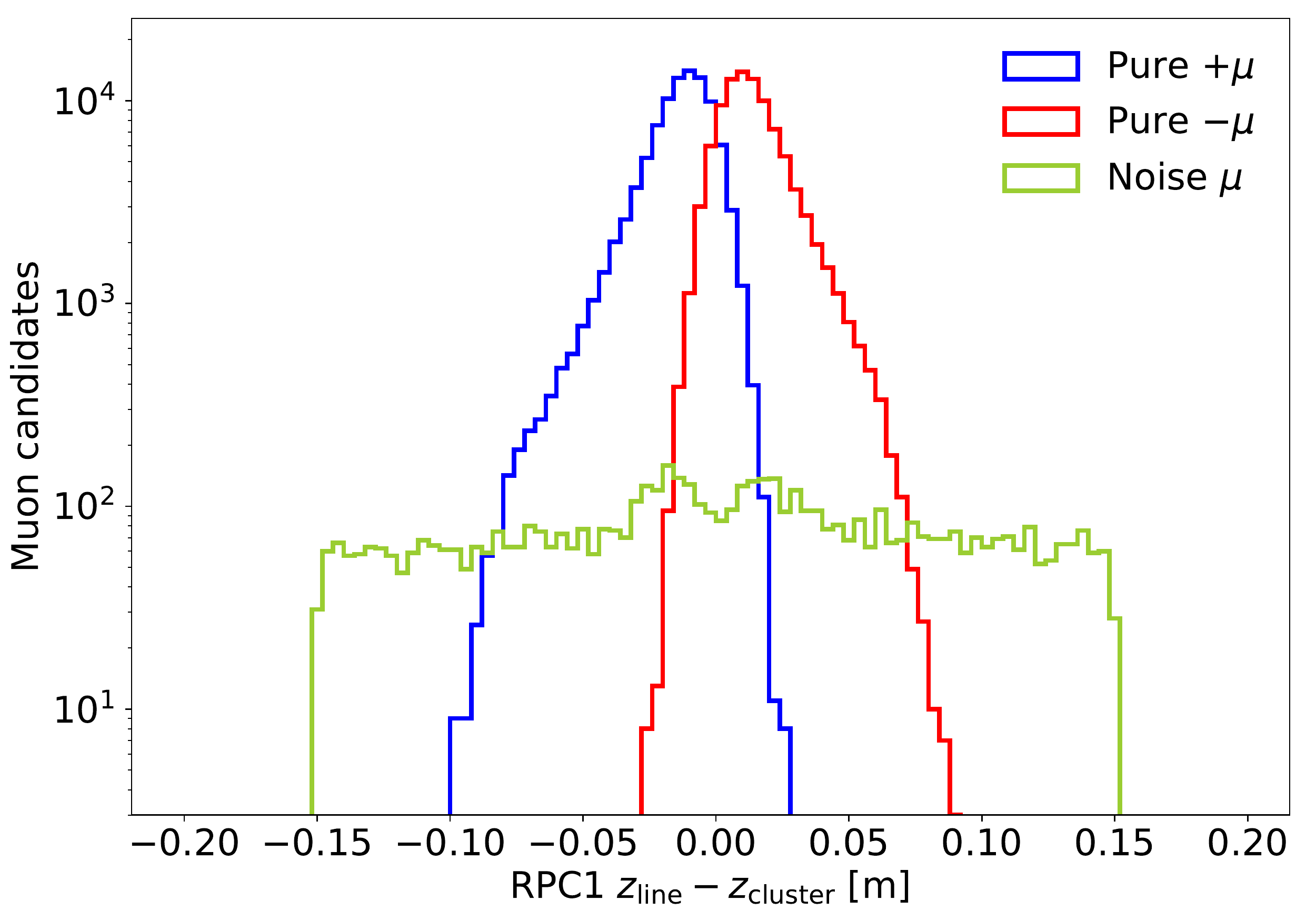}
    \includegraphics[width=0.49\textwidth]{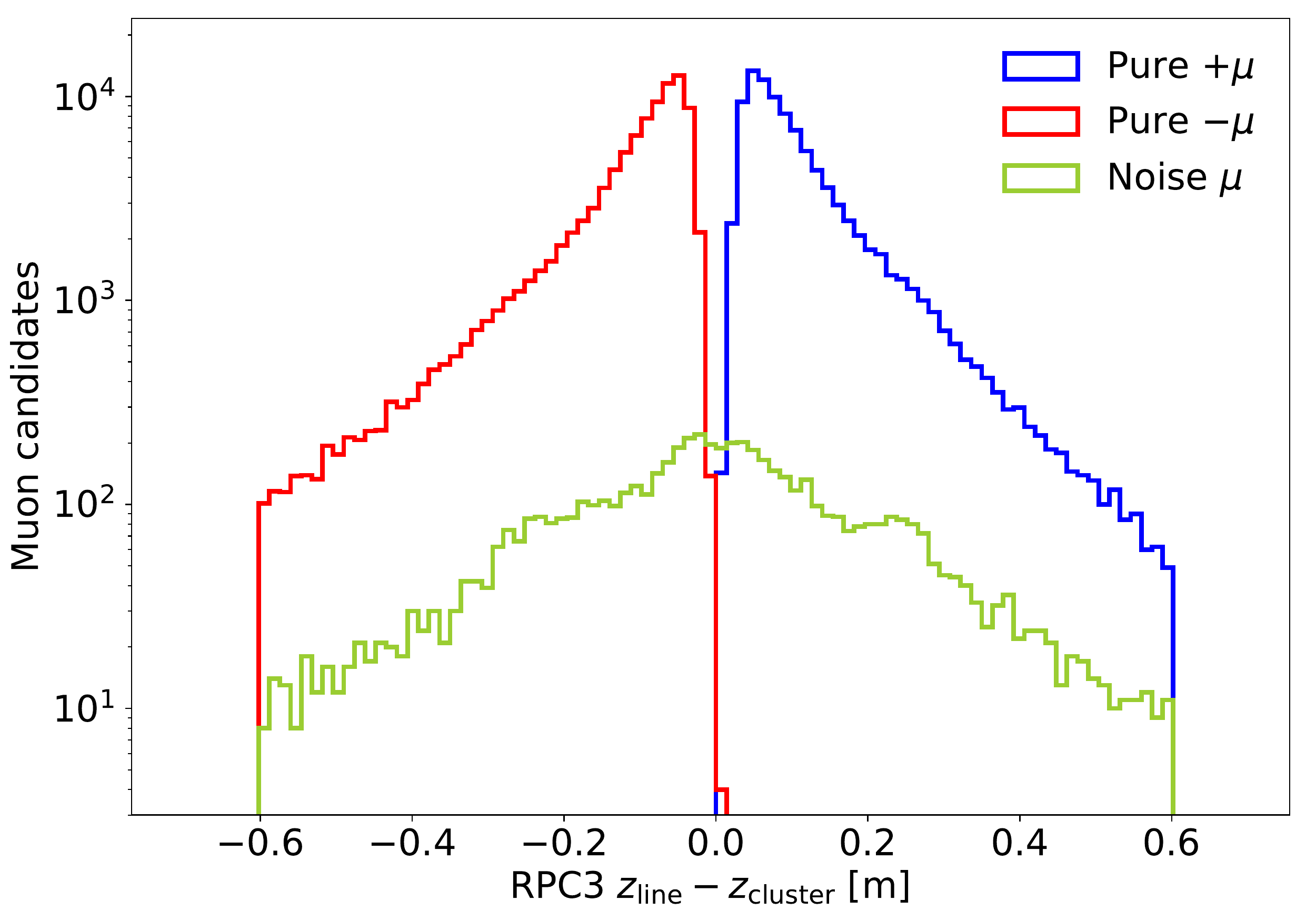}
    \caption{Differences of $z$ coordinates between the impact point of the seed line and the cluster position in the RPC1 (top) and RPC3 (bottom) layers. Positively charged muon candidates are shown in blue and negatively charged muon are shown in red. Muon candidates that include a noise cluster are shown in green.}
        \label{fig:diff}
\end{figure}

\begin{figure}[ht]
\centering
\includegraphics[width=0.49\textwidth]{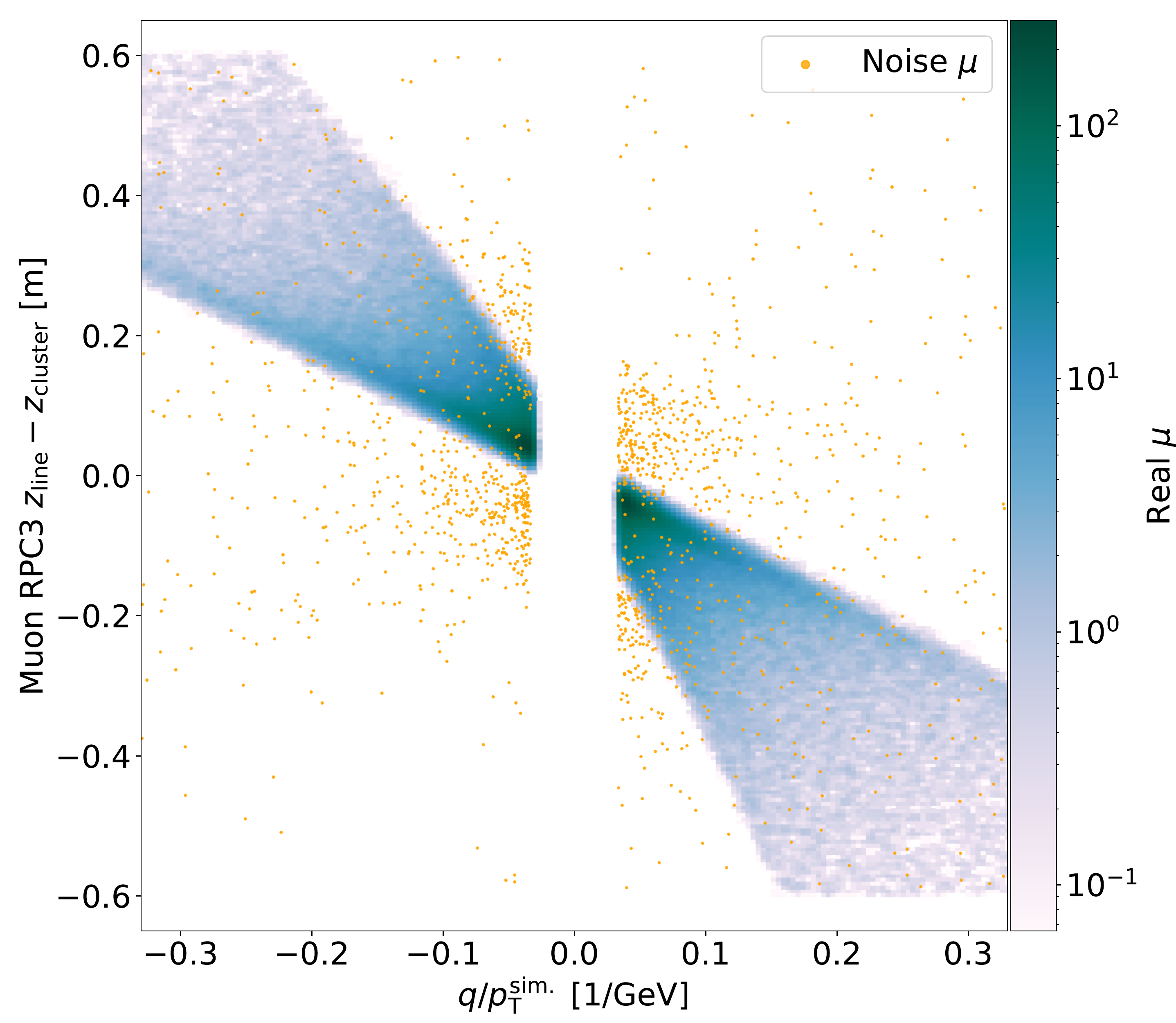}
\caption{Differences of $z$ coordinates between the impact point of the seed line and the cluster position in the RPC3 layer plotted as a function of simulated muon $1/\pT$. Colour coded density map shows candidates with only muon hits. Orange dots show candidates that include at least one noise cluster.}
    \label{fig:diff2d}
\end{figure}

Coordinates of the reconstructed clusters are used to compute three input variables for the neural network regression model. The $z$-coordinate of the RPC2 seed cluster is the first input variable. In RPC1 and RPC3 layers, the $z$ coordinate differences between the impact point of the seed line and the cluster position serve as other two input variables. These two variables are shown in Figure~\ref{fig:diff}, separately for muon candidates without noise clusters (pure muons), and for muon candidates with a noise cluster (noise muons). Figure~\ref{fig:diff2d} plots these variables as a function of the simulated muon $1/\pT$ for muon candidates without noise clusters, and for muon candidates with a noise cluster. There is a clearly visible linear relation between the $z$ differences and $1/\pT$ for the pure muon candidates. For the noise muon candidates, there are no strong correlations due to the randomness of noise hit positions.

The three input variables are linearly transformed to produce distributions with the same mean and standard deviation values in order to improve regression model performance. The inverse of the simulated muon transverse momentum times its charge ($q/\pT$) is used as the regression target. The network output serves as a prediction of muon $q/pT$ value. The $q/\pT$ target improves numerical stability of the neural network parameter optimisation because deflections of muons with $\pT > 20$~GeV are of the order of the strip width and decrease further with higher $\pT$ values. Therefore, the network cannot distinguish positively charged and negatively charged muons with $\pT>20$~GeV.

\begin{figure}[ht]
    \centering
    \includegraphics[width=0.49\textwidth]{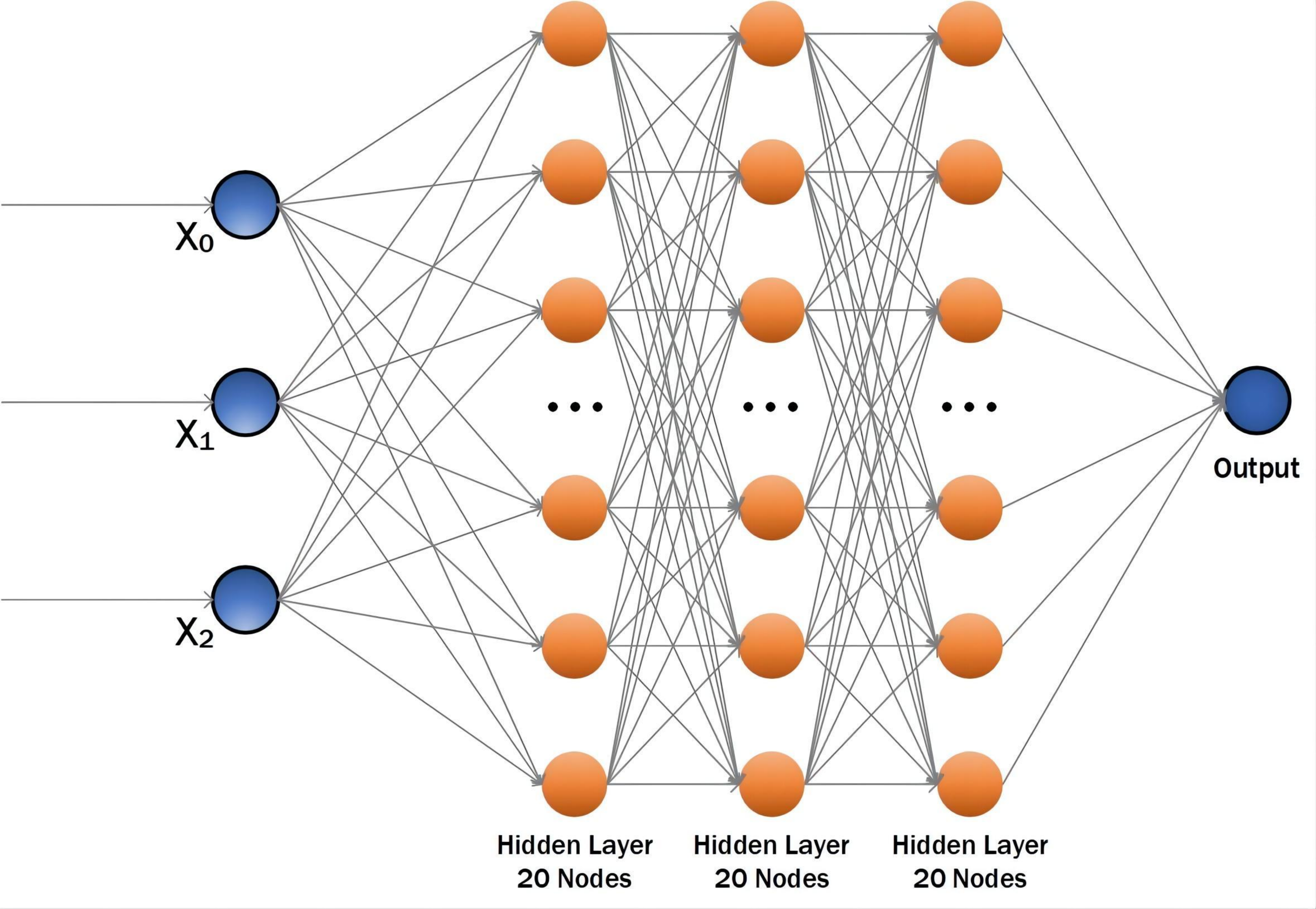}
    \caption{Structure of the neural network model.}
        \label{fig:nn}
\end{figure}

A neural network model is trained using the PyTorch library with the linear loss function, which is defined as the mean of absolute differences between simulated and predicted $q/\pT$ values. Statistically independent samples were used for training and testing different network configurations. Only muon candidates without noise clusters were used in the training. This approach was found to improve network performance and speed up the training convergence.

Several networks with different numbers of neuron nodes per layer were tested and no strong dependence on the network size was observed for configurations with more than 20 nodes per layer. The selected network configuration contains three fully connected hidden layers with 20 nodes in each layer, as illustrated in Figure~\ref{fig:nn}. The network nodes use the rectified linear unit (ReLU) activation functions. Other activation functions, such as sigmoid and hyperbolic tangent, were also tested. All of the tested activation functions result in similar performance for predicting muon $q/\pT$. Thus, the ReLU function was selected because it can be easily implemented in FPGA logic. The other activation functions can increase latency and may require more logic resources~\cite{afunc}.

\section{Neural network performance}
\label{sec:result}

\begin{figure}
    \centering
    \includegraphics[width=0.49\textwidth]{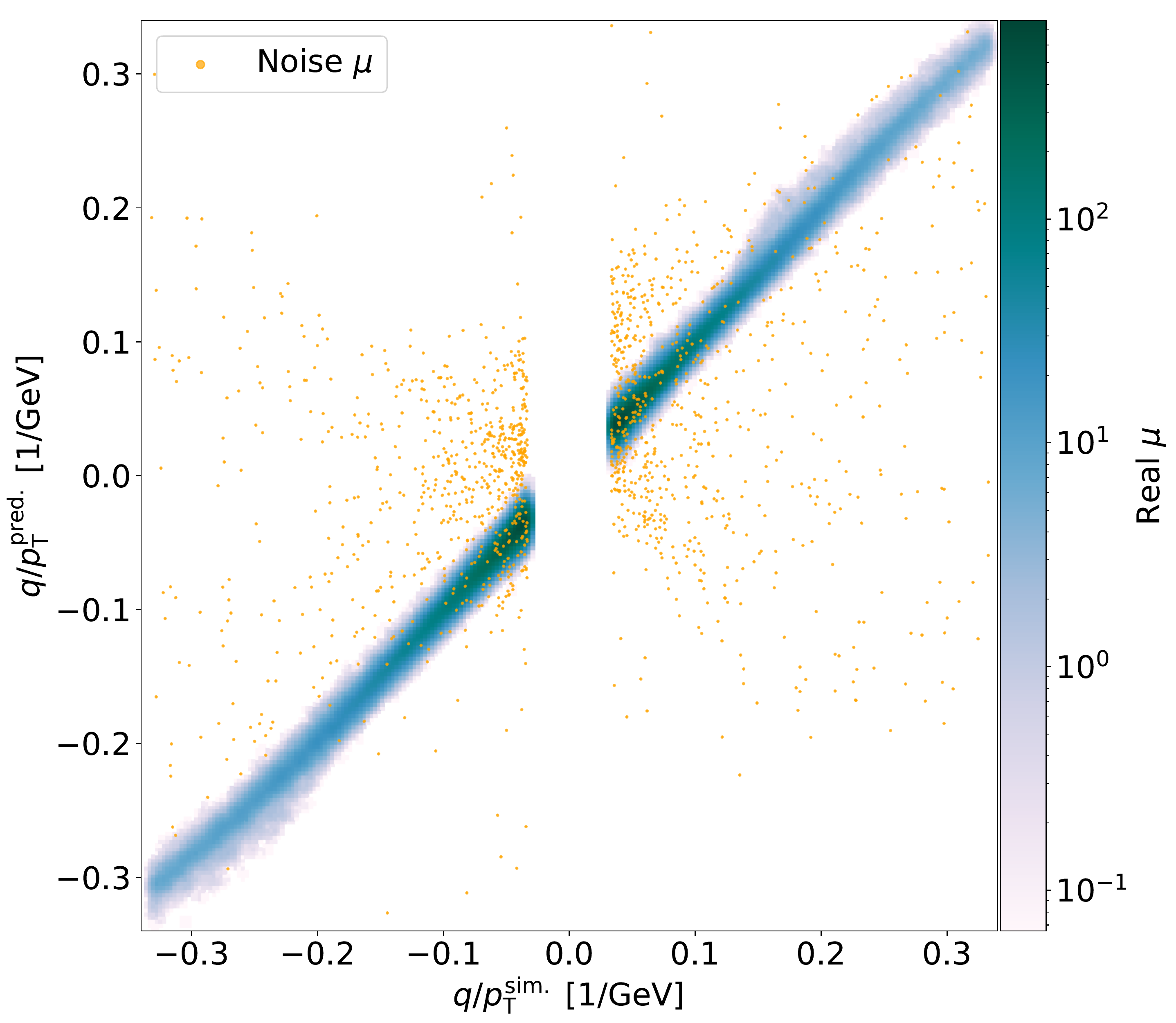}
    \caption{Output of the neural network plotted as a function of the true simulated muon $q/\pT$ using network trained with events including only muon hits. Colour coded density map shows muon candidates without any noise hits. Orange dots show muon candidates that include at least one noise cluster.}
        \label{fig:out1}
    \end{figure}

\begin{figure}
    \centering
    \includegraphics[width=0.49\textwidth]{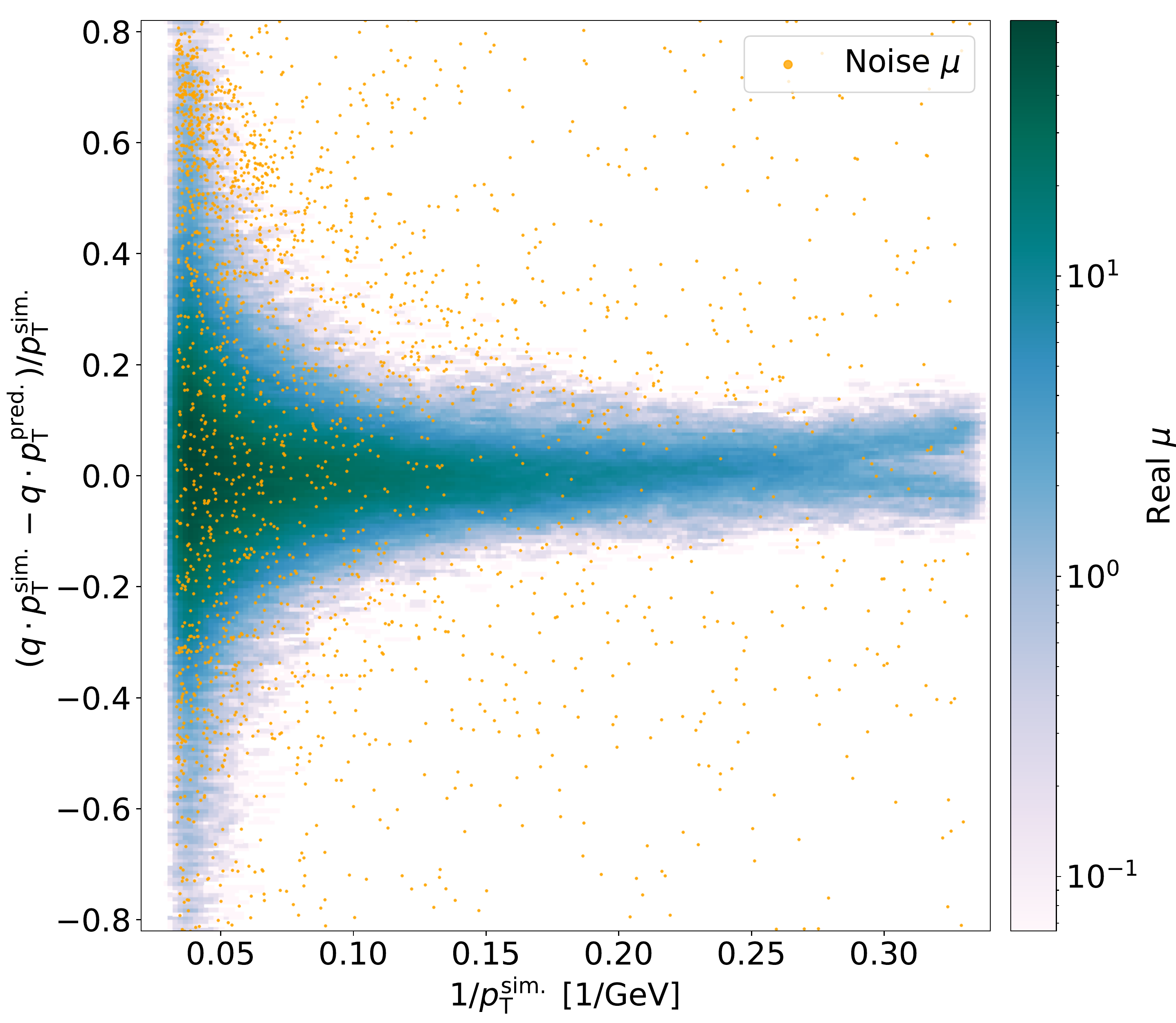}
    \caption{Relative differences between predicted and true muon $q \cdot \pT$ plotted as a function of true simulated muon $1/\pT$ using network trained with muon candidates without noise hits. Colour coded density map shows muon candidates without any noise hits. Orange dots show muon candidates that include at least one noise cluster.}
    \label{fig:resol}
    \end{figure}

Figure~\ref{fig:out1} shows the muon $q/\pT$ values predicted by the neural network plotted as a function of the true simulated  $q/\pT$, where muon candidates with and without noise hits are shown separately. The relative differences between simulated and predicted muon $q\cdot \pT$ values are plotted as a function of the simulated muon $\pT$ in Figure~\ref{fig:resol}. These two figures show that our regression model accurately predicts muon $q/\pT$ values for muon $\pT<20$~GeV. As expected, the $\pT$ resolution is best at low $\pT$ values and then slowly degrades with increasing $\pT$ because muon displacements due to the magnetic field become comparable to the strip width for muon $\pT>20$~GeV.

\begin{figure}
    \centering
    \includegraphics[width=0.49\textwidth]{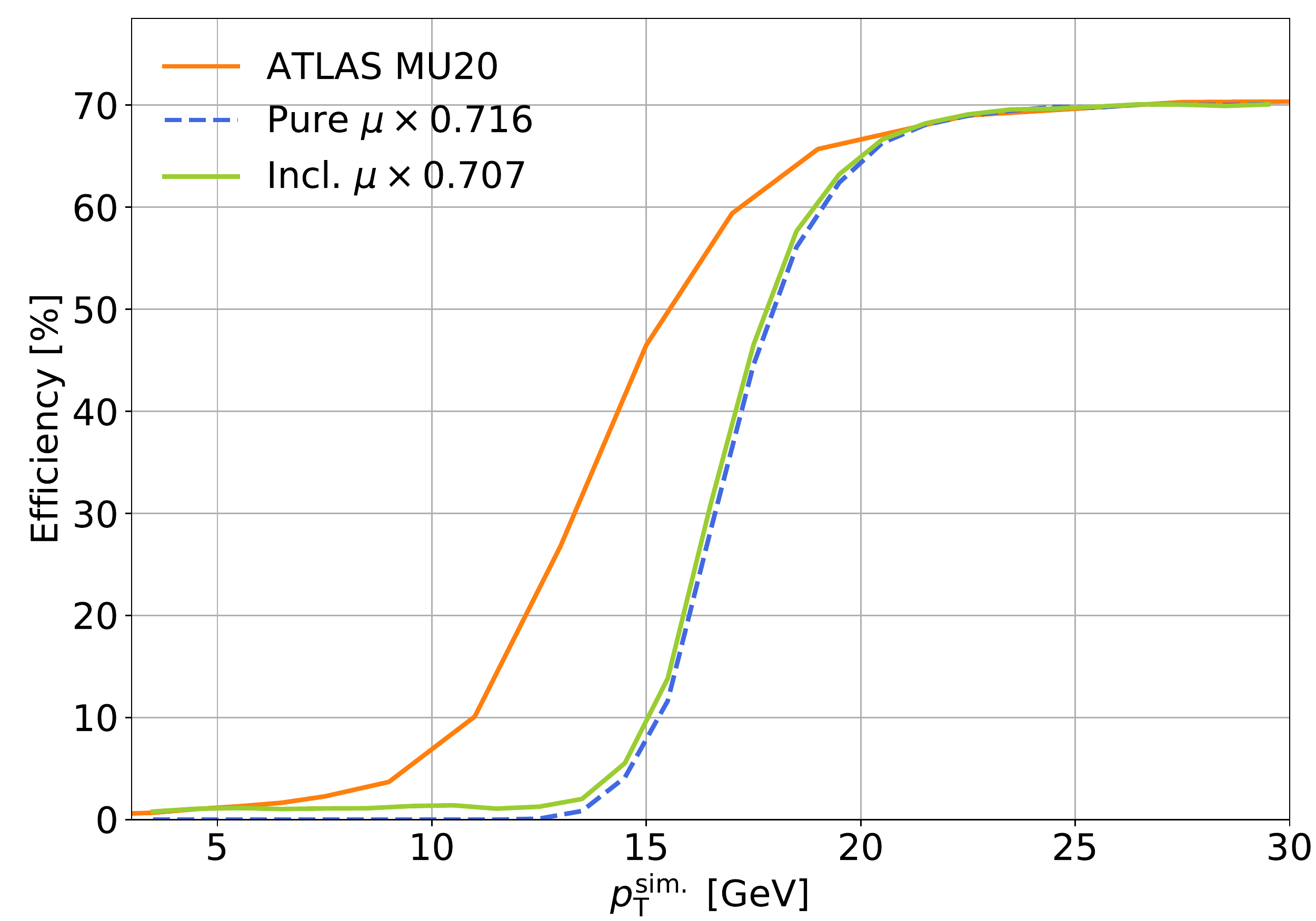}
    \caption{Efficiency of selecting muon candidates plotted as a function of the muon {\pT}. Shown in orange is the ATLAS data efficiency for the MU20 trigger threshold of the present RPC detector~\cite{RPCperf}. Shown in green (blue) is efficiency for selecting simulated muons with (without) noise hits. These two curves are scaled to obtain the same efficiency as the MU20 trigger for $\pT>25$~GeV.}
    \label{fig:eff1}
    \end{figure}

Performance of the neural network regression model was evaluated by computing the efficiency to select muon candidates with $\pT>20$~GeV. Figure~\ref{fig:eff1} shows the efficiency of selecting muons with $\pT>20$~GeV for muon candidates without noise clusters and for inclusive muon candidates that also include candidates with noise clusters. These two efficiency curves are compared to the reference curve. This curve corresponds to the efficiency of the RPC MU20 trigger which selects muons with $\pT>20$~GeV in the barrel detector region. This reference efficiency~\cite{RPCperf} was measured using collision data recorded by the ATLAS detector in 2018. The efficiency of the MU20 trigger reaches the plateau at 70\% due to the RPC detector coverage gaps because of inefficient modules and presence of support structures~\cite{RPCperf}. Since the present simulation model does not include these effects, the efficiencies predicted by the regression model are scaled to the same value as the MU20 trigger efficiency curve for $\pT>25$~GeV, in order to allow easier comparisons of efficiency curve shapes.

    \begin{figure}
        \centering
        \includegraphics[width=0.49\textwidth]{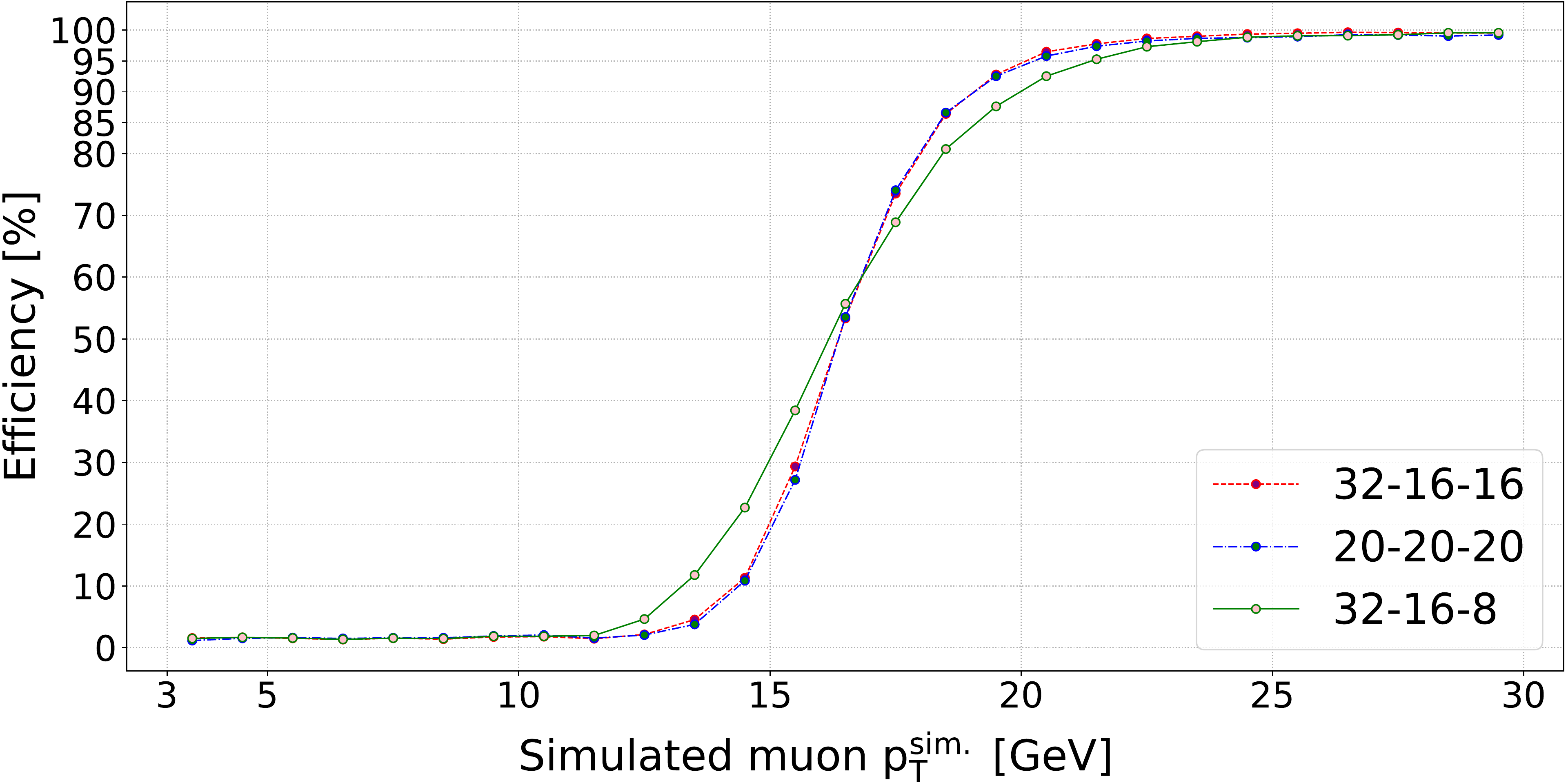}
        \caption{Efficiency of selecting muon candidates plotted as a function of the muon {\pT} for three different neural network configurations.}
        \label{fig:effnn}
        \end{figure}

As discussed previously, several network configurations were tested. Figure~\ref{fig:effnn} shows efficiency for selecting muons with $\pT>20$~GeV for three different network configurations: 32-16-16, 32-16-8, and 32-20-20, where three numbers represent a number of nodes in layers 1, 2 and 3, respectively. A smaller neuron network with 8 nodes in the last layer (32-16-8) results in the worse performance. Two larger networks produce similar results, with 20-20-20 configuration resulting in a slightly better performance. This network was therefore chosen as the final configuration.

The main purpose of the present work is to improve RPC detector resolution for measuring muon $\pT$. Achieving this goal would lead to better rejection of low-$\pT$ muon candidates which dominate the sample of collisions events accepted by the RPC muon trigger~\cite{RPCperf}. The developed regression model produces the sharper rising efficiency curve than that of the MU20 trigger. This sharper curve would lead to lower trigger acceptance of the low-$\pT$ muons with incorrectly measured $\pT$. However, the present simulation model does not fully account for all effects present in the real detector. For example, there are regions in the ATLAS barrel muon spectrometer where the magnetic field is not uniform or is weaker than what is assumed by our model. Therefore, this potentially better performance  of the regression model will be verified in the follow up work using a more precise simulation model and collision data.

\section{FPGA implementation}
\label{sec:fpga}

This section presents details of the implementation of the neural network regression model in FPGA code. Section~\ref{sec:precis} motivates our choice for numerical precision of the FPGA logic implementation. Section~\ref{sec:logic} presents details of the FPGA implementation. Section~\ref{sec:fpgaperf} reports measurements of FPGA timing performance, latency, and resource utilisation. Finally, Section~\ref{sec:fpgatests} presents results of testing the FPGA logic implementation using simulated events.

\subsection{Data precision for FPGA logic implementation}
\label{sec:precis}

The neural network model was implemented in FPGA using 16-bit binary fixed-point numbers. This choice results in fast network execution speed and low usage of FPGA resources.  Prior to the FPGA implementation, the number of bits allocated to integer and decimal (fractional) parts of fixed-point 16-bit numbers were carefully studied. This study was motivated by the requirement of avoiding frequent integer overflows while maintaining sufficient decimal accuracy for 16-bit arithmetic operations. 

The output values of neuron activation functions of the hidden layers typically range between -8 and 8. To cover nearly all of neuron output values, at least 5-bit integer part is required for the fixed point calculations. One or two extra bits can be reserved to avoid overflows for addition operations. Therefore, only 9 to 11 bits can be used for the decimal part.

\begin{figure}[ht]
	\centering
	\includegraphics[width=0.99\linewidth]{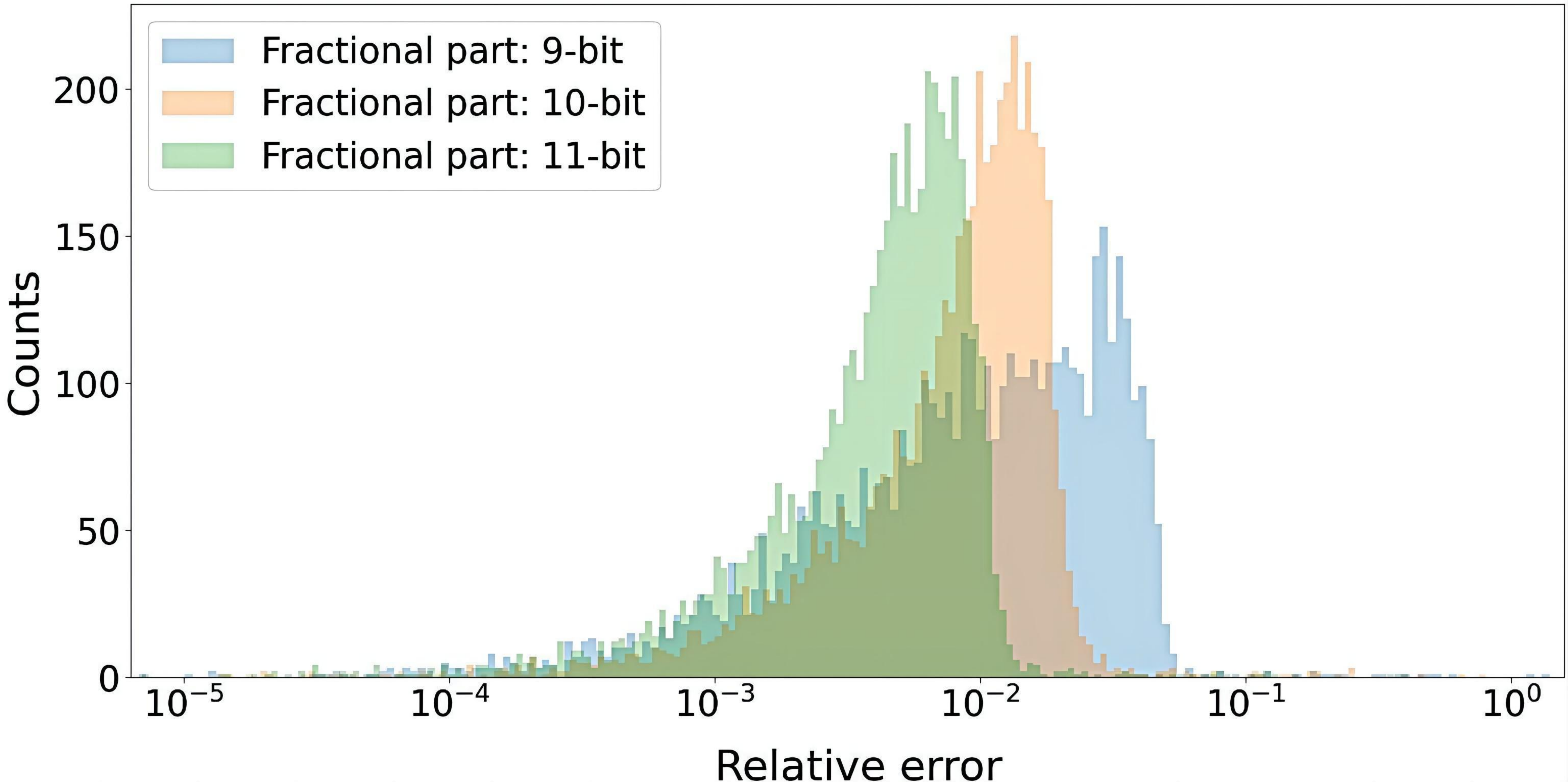}
    \caption{Errors for network output values computed as the relative differences between values obtained using the implementation with 16-bit precision and implementation with 32-bit precision. Three different results with the decimal precision of 9, 10 and 11 bits are compared in the plot.}
	\label{fig:numbs}
\end{figure}

To select the best possible decimal precision, network output values were computed using the network implemented in the custom Python code with three different values of the decimal precision: 9, 10 and 11 bits. These network output values were compared with the values obtained using the same implementation but using 32-bit floating-point precision. Relative errors between each of the three calculations and the 32-bit implementation are shown in Figure~\ref{fig:numbs}. The implementation using 10-bit decimal precision achieves sufficient accuracy so this solution was adopted in order to reserve one extra bit for the integer part. This choice helps to further reduce integer overflows for 16-bit arithmetic operations.

\subsection{FPGA logic design} 
\label{sec:logic}

\begin{figure}[ht]
    \centering
    \includegraphics[width=0.99\linewidth]{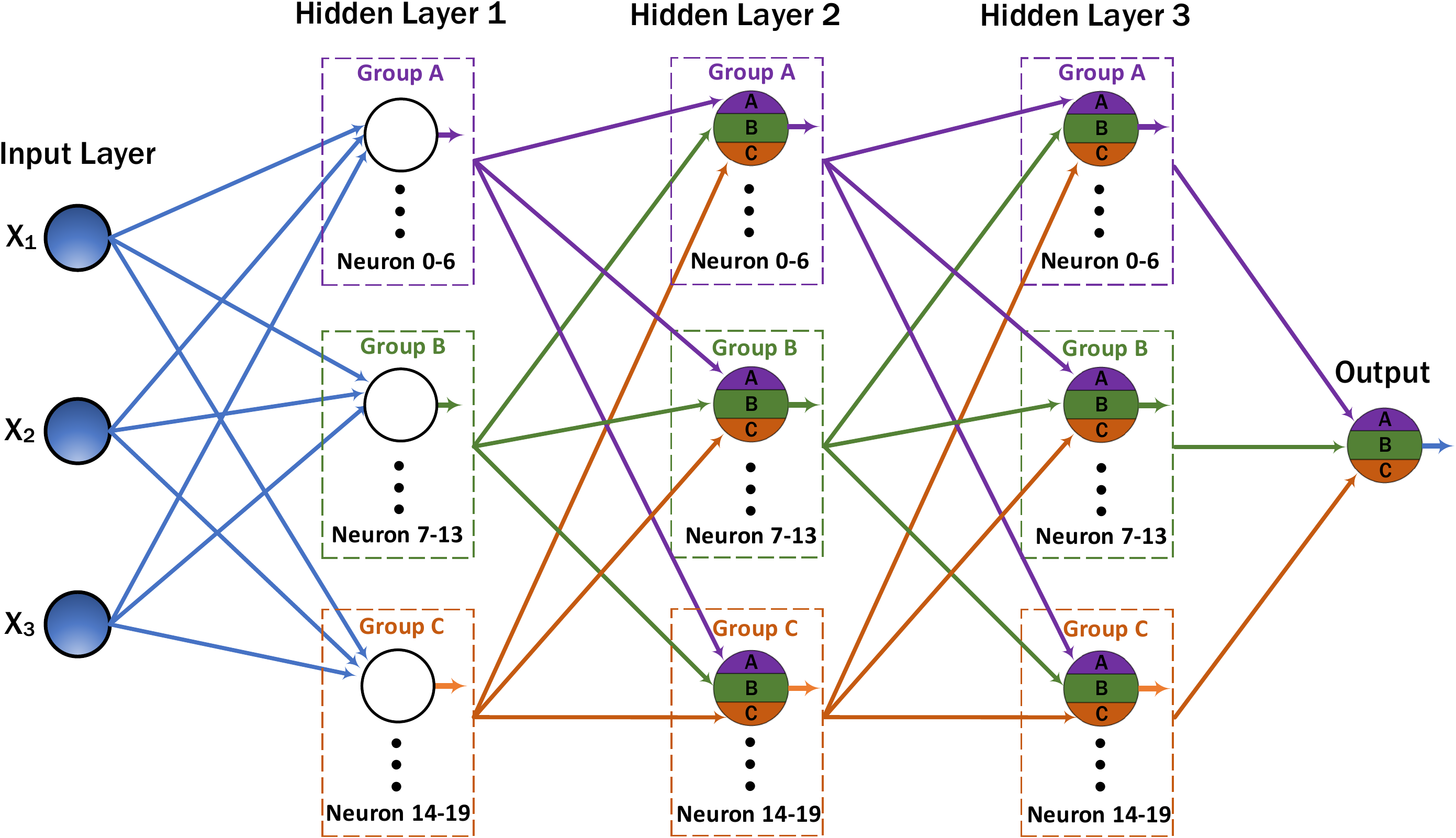}
\caption{FPGA data flow of the neural network logic. Output neuron and each neuron in layers 2 and 3 contain three PEs, denoted A, B, and C. Each PE receives and processes data from the corresponding group in the previous layer.}
\label{fig:nnflow}
\end{figure}

The full neural network was implemented in the FPGA code using HDL~\footnote{Source code is available at \href{https://github.com/rustemos/hdl4nn}{https://github.com/rustemos/hdl4nn}}. Each layer of this fully connected network represents the multiplication of an input data vector and a matrix containing layer weights. Each neuron performs the inner product of two vectors, one vector representing input data and another representing specific node weights. The FPGA implementation of these operations is built using three main logical blocks that correspond to the three neural network layers. Figure~\ref{fig:nnflow} shows the data flow diagram of our neural network implementation. Neuron layers are divided into neuron groups, each containing either 6 or 7 neurons. Each neuron group sends its data serially and simultaneously with other groups in the same layer. In layers 2 and 3, each neuron receives data from all neurons in the previous layers. This data is transferred in parallel by three neuron groups with the receiving neuron processing that data in parallel using three processing elements (PEs). This division of the neurons into three groups reduces the layer latency by a factor of three. The same logic is also used for the output neuron.

\begin{figure}[ht]
    \centering
    \includegraphics[width=0.99\linewidth]{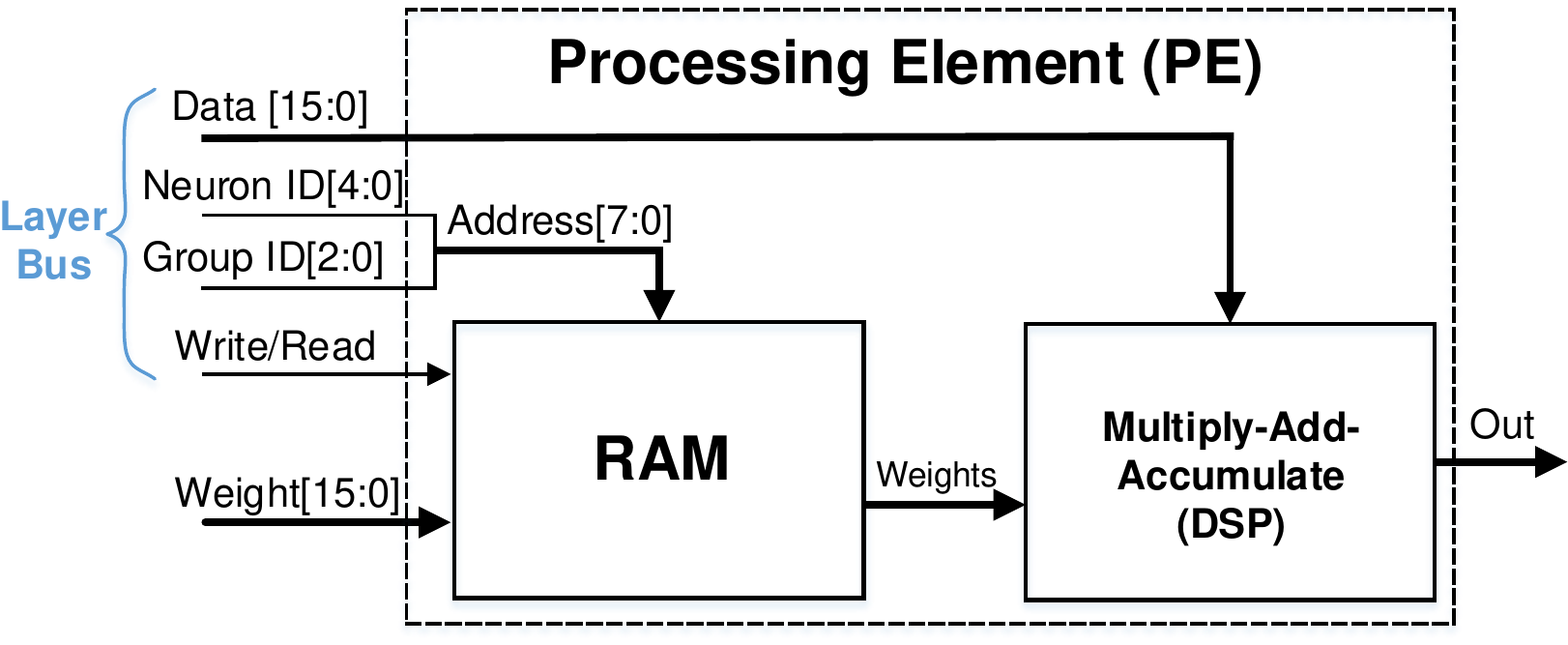}
\caption{FPGA logic design for one processing element (PE). Random access memory unit (RAM) stores node weights. Multiply-add-accumulate unit (MAC) implements arithmetic operations using a single DSP.}
\label{fig:pe}
\end{figure}

The PE is the basic unit for FPGA implementation of the neural network logic. It contains two logical units: a multiply-add-accumulate (MAC) unit and a random-access memory (RAM) unit, as illustrated in Figure~\ref{fig:pe}. Input data is processed serially by each PE in order to simplify the circuit design. The MAC unit performs vector multiplication operations and the RAM unit stores weights using distributed FPGA RAM. One digital signal processor (DSP) is used by each MAC unit to perform addition and multiplication operations. In hidden layers, additional logic to perform the ReLU functionality is implemented using one DSP. In order to maximise processing speed and to reduce latency, our design does not use lookup tables and block memory.

\begin{figure*}[h]
    \centering
    \includegraphics[width=0.85\textwidth]{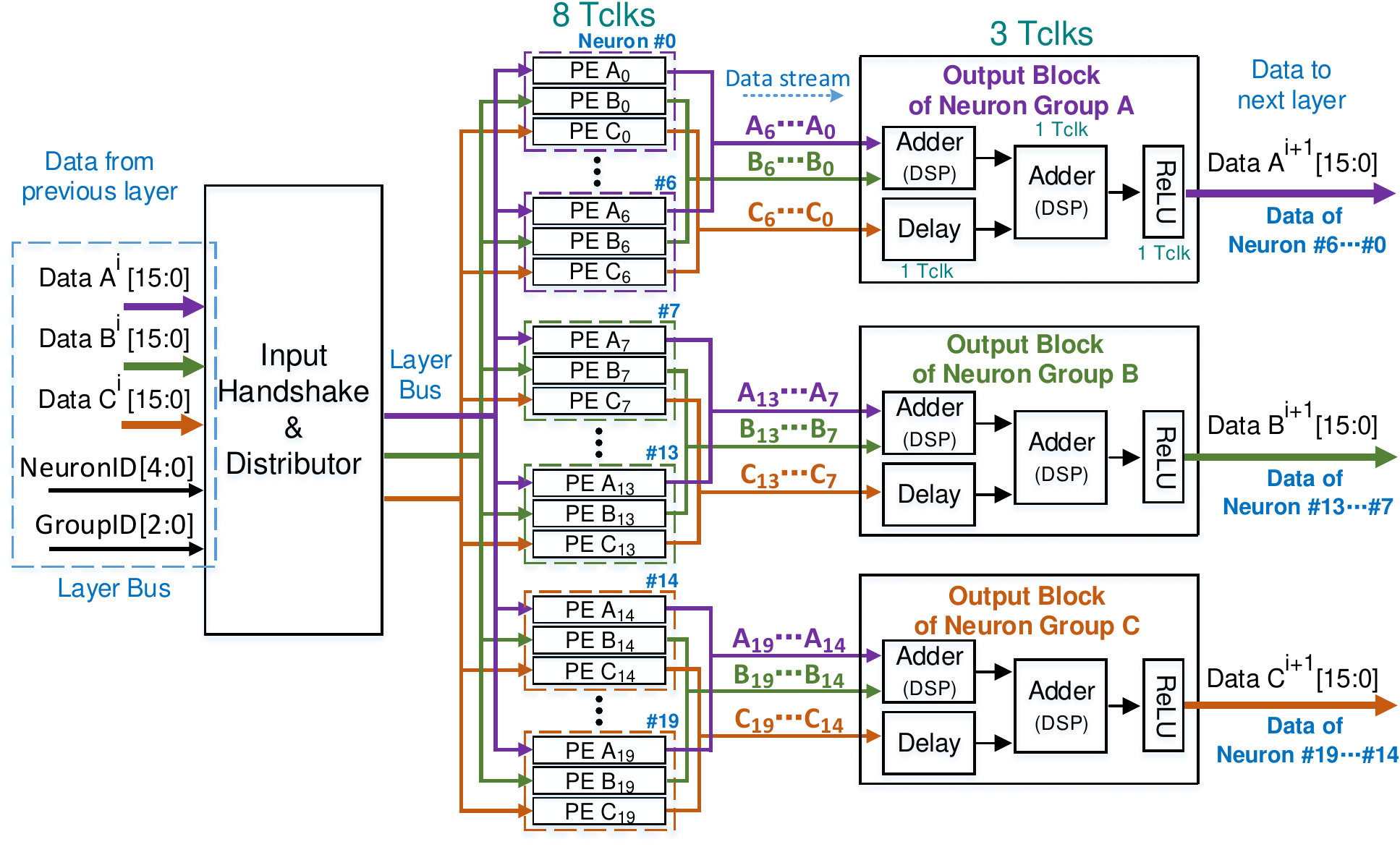}
\caption{FPGA logic design for one neuron layer.}
\label{fig:layer}
\end{figure*}

The overall structure of the hidden layer implementation is shown in Figure~\ref{fig:layer}. It consists of one input block, twenty neuron blocks, and three output blocks. Data is sent and received when a layer is ready for processing. This ready signal is propagated using dedicated handshake lines between layers. Each layer includes a distributor unit which makes 20 copies of every input data element and transmits them in parallel to PEs. Layer results are stored in the output buffer until a next layer is ready to receive data.  

The layer input block receives three data streams from three neuron groups of the previous layer (labelled A, B and C). In each data stream, output data of the seven (or six) neurons is sent serially, with one transmission per clock cycle. Each neuron receives three input data streams in parallel, with one dedicated PE performing arithmetic operations for each data stream. Arithmetic operations of every neuron node are performed in parallel with other nodes since the layer nodes are independent from each other. This parallel design reuses the same basic PE structure and leads to lower latency.

A PE processes its input data within eight clock cycles. One cycle is used by the data preparation step. Seven cycles are used by the neural node logical operations. Results of each neuron group (A, B or C) are organized into a new data stream which is sent serially to a next layer via the group's output block which functions as a pipeline. The output block sums together outputs of three PEs of each neuron and sends the resulting sum to the ReLU block which prepares a new data stream for a next layer. Two DSPs perform summation operations of the output block within two clock cycles. ReLU calculation takes one clock cycle.

In addition to the FPGA implementation described so far, two other implementations were developed for comparison purposes and to better understand the impact of design choices on resource utilisation and latency. The same clock frequency of 320~MHz and the same 16 bit precision were used for all three implementations. The first of these additional implementations used a simpler HDL design with only one PE in each neural node. This implementation is referred to as the simplified HDL design. This simpler design was in fact implemented first and then it was optimised to improve timing performance to produce our final design, at a price of somewhat larger resource utilisation. This final design is described earlier in this section and it is referred to as the optimised HDL design.

The third FPGA implementation of our model was performed using the $hls4ml$ project of Ref.~\cite{Duarte:2018ite}. The $hls4ml$ tools translate neuron network models into HLS code for FPGA implementation. It requires significantly less effort to implement multilayer networks or convolutional networks using $hls4ml$, compared to manually developing a FPGA project using HDL. Our neural network model was transformed into a format compatible with $hls4ml$. This model was then converted into an HLS project using $hls4ml$. FPGA firmware implementation was then generated using Vitis HLS. The $hls4ml$ project allows fine tuning resulting FPGA implementation by changing the so-called "reuse" factor, which determines how many multiplications are performed in one single clock period by each DSP. A higher reuse factor uses fewer FPGA resources at a price of higher latency and deadtime. Three different reuse factors were tested: 1, 2 and 4, to produce results reported in the next section.

\subsection{Performance of FPGA logic implementation}
\label{sec:fpgaperf}

\begin{figure*}[ht]
    \centering
    \includegraphics[width=0.99\textwidth]{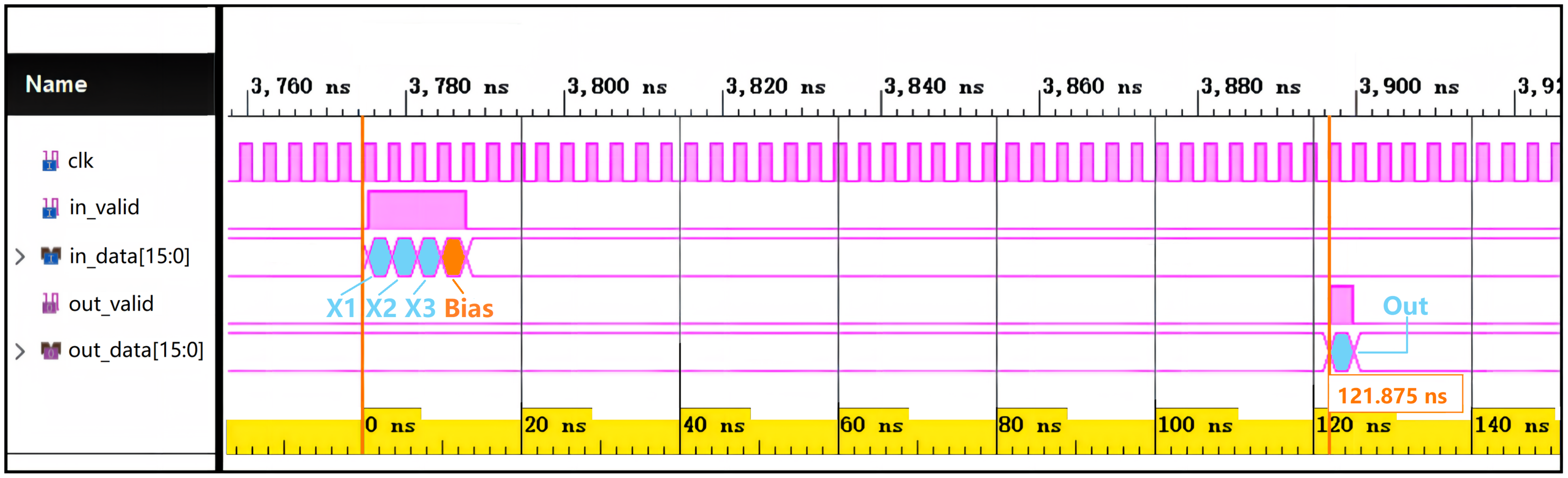}
\caption{FPGA timing simulation of the full neural network.}
\label{fig:wave}
\end{figure*}

The full network logic was simulated for the Xilinx XCKU060 FPGA using 320~MHz clock. The latency and deadtime of our optimised HDL implementaion are summarised in Table~\ref{tab:lat}, separately for each network layer. The hidden layer deatime is eight clock cycles, driven by the PE deadtime which has the maximum deadtime among all layer blocks. The input layer contains fewer PEs than the hidden layer because of the smaller input data size. The output layer contains just one neuron without the ReLU logic. As a result, these two layers have smaller latency and deadtime values. The timing simulation of the full network is shown in Figure~\ref{fig:wave}. The latency and deadtime of the optimised HDL implementation of the full network are $39\times 3.125~\text{ns} = 121.875~\text{ns}$ and $8\times 3.125~\text{ns} = 25~\text{ns}$, respectively. The total network latency is the sum of latencies of individual components. The overall network deadtime is determined by the maximum deadtime of individual components.

\begin{table}
    \centering
    \begin{tabular}{c|c|c|c}
    \hline
    & \thead{Deadtime \\ (cycles)} & \thead{Latency\\ (cycles)} & DSPs \\
    \hline
    Layer 1 & 5 &  7 & 20\\
    \hline
    Layer 2 & 8 & 11 & 66\\
    \hline
    Layer 3 & 8 & 11 & 66\\
    \hline
    Output layer & 8 & 10 & 5\\
    \hline
    \end{tabular}
    \caption{Latency and deadtime of the individual layers of the optimised HDL design}
    \label{tab:lat}
\end{table}

The latency, deadtime, and resource utilisation are summarised in Table~\ref{tab:resource} for all three FPGA implementations. For the optimised HDL design, the utilisation of the available DSP resources is less than 6\% and the utilisation of the available LUTs is less than 1.5\%. For comparison, the fastest HLS design (with reuse factor set to one) uses significantly larger fractions of logic resources, with comparable latency performance. These results demonstrate that carefully tuned HDL designs can produce resource-efficient implementations while maintaining sufficiently fast timing performance. In particular, our optimised HDL design of the full neural network was realised in FPGA hardware with efficient use of its resources, and with low latency and deadtime parameters. This result opens possibilities for real-time machine learning applications in the future ATLAS trigger system.

\begin{table*}
    \centering
    \begin{tabular}{c|c|c|c|c|c}
    \hline
    $ $ & Deadtime (cycles) & Latency (cycles) & DSPs & LUTs & Registers \\
    \Xhline{2.0\arrayrulewidth}
    \textbf{Optimised HDL design}  & 8 & 39  & 157 (5.7\%)  & 4659 (1.4\%) & 7370 (1.1\%) \\
    \hline
    \textbf{HLS (reuse = 1)}       & 2 & 23  & 677 (24.5\%) & 30952 (9.2\%) & 15507 (2.3\%) \\
    \Xhline{2.0\arrayrulewidth}
    HLS (reuse = 2)       & 6 & 94  & 430 (15.6\%) & 38701 (11.5\%) & 8916 (1.4\%) \\
    \hline
    HLS (reuse = 4)       & 8 & 100 & 225 (8.2\%)  & 35972 (10.7\%) & 6301 (1.0\%) \\
    \hline
    Simplified HDL design & 24 & 98 &  65 (2.4\%)  & 9949 (3.0\%) & 10257 (1.6\%)\\
    \hline
    \end{tabular}
    \caption{FPGA resource utilization, deadtime and latency for the neural network implementations using our optimised and simplified HDL designs, and using the HLS design obtained using the $hls4ml$ project. Resource utilisation fractions (shown in percent) are computed for the Xilinx XCKU060 device.}
    \label{tab:resource}
\end{table*}

\subsection{Test results of FPGA logic implementation}
\label{sec:fpgatests}

After the HDL design was finished, the full network circuit has been tested using simulation. Simulated muon events were generated and processed using the methods detailed in Section~\ref{sec:design}. A simulation test project was developed using Questa Advanced Simulator and SystemVerilog language. An interface was designed for communications between software and Verilog blocks. Input data for network evaluation is sent to the Verilog block and network output data is read back through this interface. The generated dataset was adapted for interfacing with the input ports of the FPGA-based neural network.

Candidate muons were evaluated using the FPGA simulation and neural network results produced by this simulation were compared with those obtained using the Python implementation. In total, 200 thousand simulated events were generated and processed, with $\pT$ values uniformly distributed between 3~GeV ad 30~GeV and the muon angle uniformly distributed between 40 to 85 degrees. The inference result of the FPGA neural network simulation (i.e. predicted value of $q/\pT$) is compared event by event with the Python result. 

\begin{figure}[ht]
	\centering
	\includegraphics[width=0.99\linewidth]{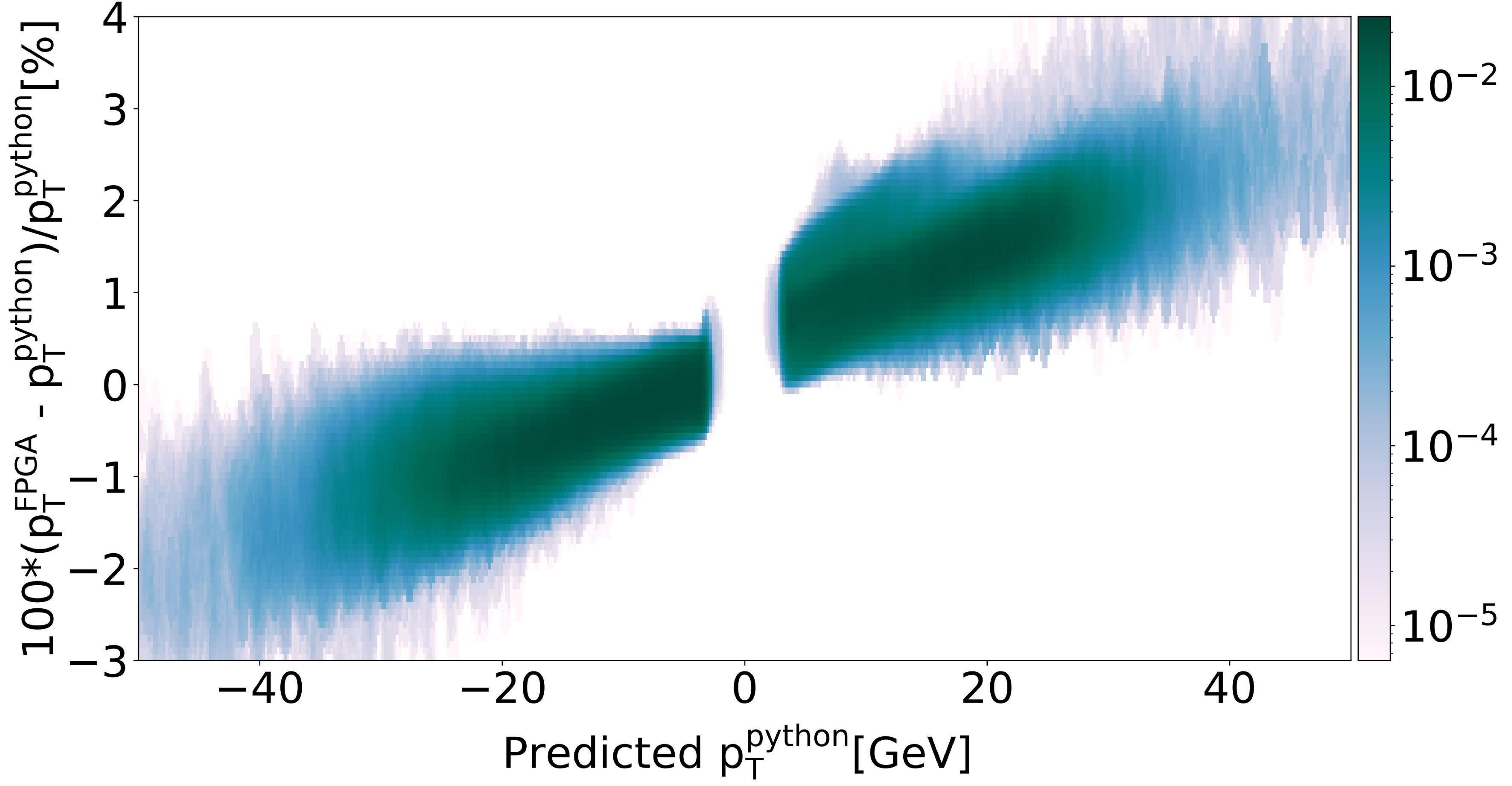}
    \caption{Relative differences between $\pT$ values computed by the software and FPGA plotted as a function of $\pT$ predicted by the software.}
	\label{fig:simtestpt}
\end{figure}

Figure~\ref{fig:simtestpt} shows the correlations of the results of the FPGA simulation with the Python results. The $\pT$ values predicted by the FPGA simulation are nearly the same as those predicted by the Python implementation for predicted $\pT < 100$~GeV. The output $\pT$ values reach up to and above 50~GeV due to larger errors in measuring a muon transverse momentum for $\pT > 20$~GeV. This is expected since the RPC detector was designed to identify muons with $\pT$ up to 20~GeV. Above this threshold, the muon curvature becomes comparable to the RPC strip width and therefore the RPC detector cannot measure precisely $\pT$ of such muons. Therefore, these differences between the FPGA simulation and Python implementation above 20~GeV do not degrade the performance of the regression model for distinguishing muons with $\pT > 20$~GeV from those with $\pT < 20$~GeV.

\begin{figure}[ht]
	\centering
	\includegraphics[width=0.99\linewidth]{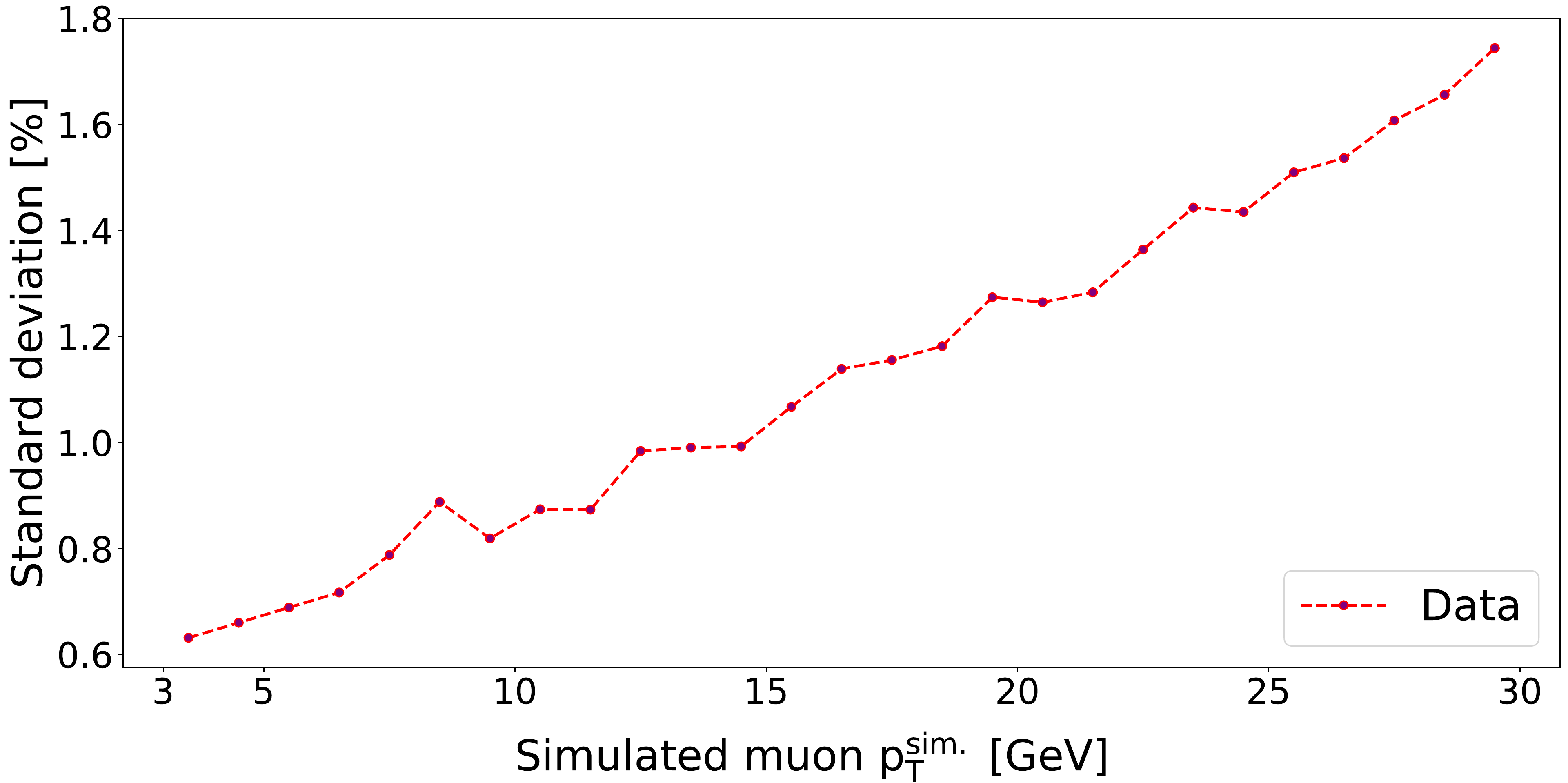}
    \caption{Standard deviation of the relative differences between $\pT$ values predicted by the python code and by FPGA simulation plotted as a function of $\pT$.}
	\label{fig:simtesterr}
\end{figure}

\begin{figure}[ht]
	\centering
	\includegraphics[width=0.99\linewidth]{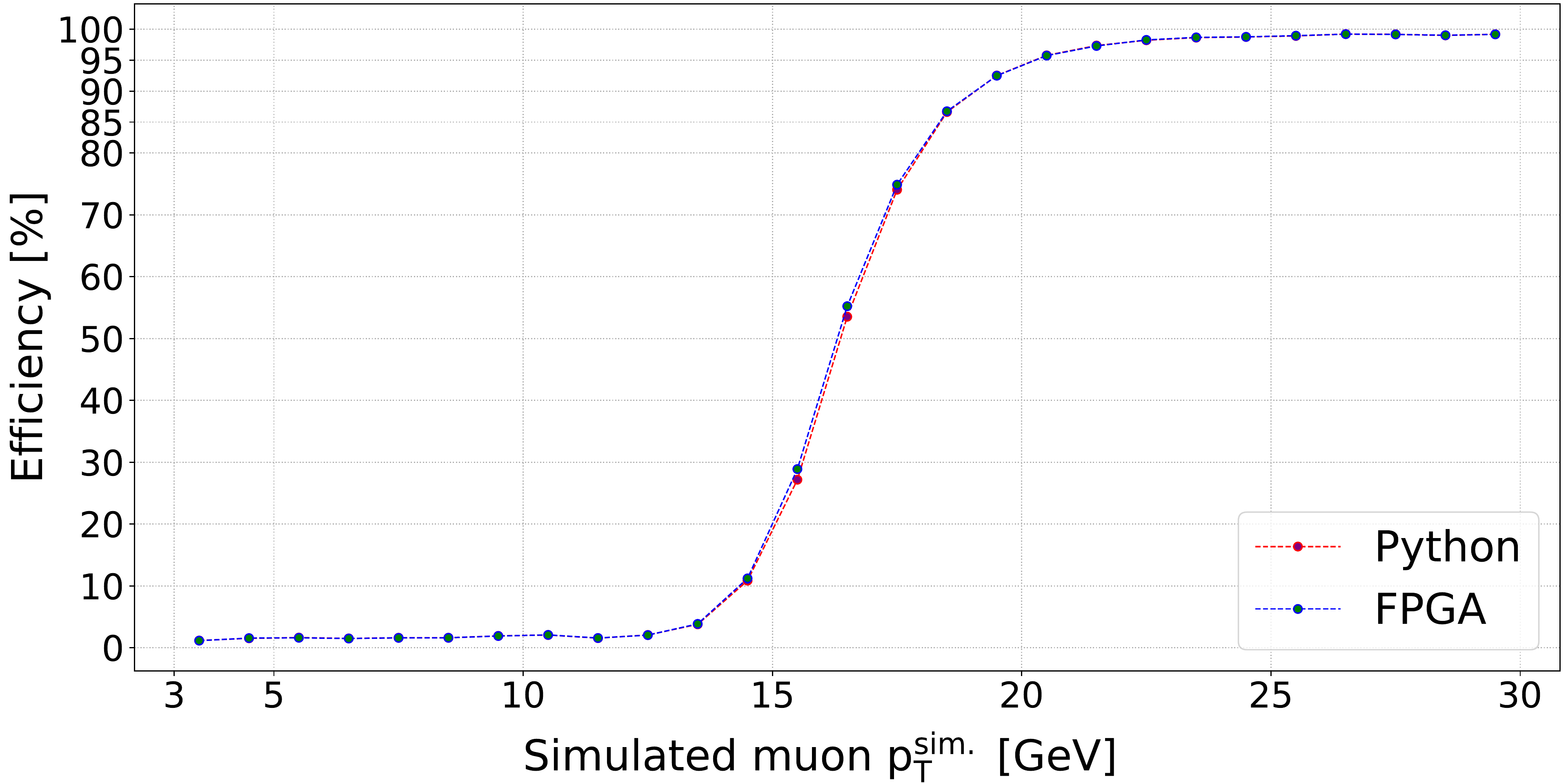}
	\caption{Muon trigger efficiency plotted as a function of the simulated muon $\pT$ for the software (red) and FPGA (blue).}
	\label{fig:simtesteff}
\end{figure}

The relative error between these two approaches is shown as a function of the muon $\pT$ in Figure~\ref{fig:simtesterr}. Within the working $\pT$ range of the RPC detector, the errors introduced by the 16-bit arithmetic of the FPGA hardware do not impact the network performance for predicting muon $\pT$. As shown in Figure \ref{fig:simtesteff}, the trigger efficiency curve computed by the FPGA simulation is consistent with the result computed by the Python program.

\subsection{Discussion of FPGA implementation}
\label{sec:fpgsum}

Verilog HDL is used in this paper to realise the neural network on the FPGA device. Another approach using HLS was also studied. The major advantage of HLS over HDL is that the HLS programming environment and tools are far more convenient for developing FPGA code. Hence, using the HLS allows implementation of complex algorithms in FPGA hardware with short development cycles. HLS implementations are also usually more flexible, for example allowing non-expert users to easily change a neural network size and numerical precision of its arithmetic operations.

The disadvantages of the HLS tools are also well-known. Circuits designed using the HLS tools often utilise more logic resources than comparable algorithms implemented using well-tuned HDL code. In addition, HLS-designed circuits often run at a lower clock frequency than a maximum available device frequency. For example, Refs.~\cite{Marjanovic:2019tle,Millon2020ACS} show that the HLS-designed circuits consume a factor of three more FPGA logic resources compared to the circuits which were manually designed using HDL. These observations are also confirmed by our work. The logic utilisation of our HDL implementation is lower compared to that of the HLS design performed using the $hls4ml$ project. This demonstrates the utility of using HDL in situations where neural network performance is constrained by scarcity of available FPGA resources. 

In the context of muon trigger upgrades at the LHC, convolutional neural networks (CNNs) were previously applied for detecting muon candidates by the upgraded ATLAS RPC detector~\cite{epjcroma}. Our approach instead focuses on measuring muon candidate $\pT$ using a neural network regression model. A simple network architecture is deployed to measure more precisely $\pT$ in order to reduce the dominant source of background events due to muons with mismeasured $\pT$. A carefully tuned HDL design is used to implement the neural network model in FPGA firmware. The resulting FPGA implementation has a lower latency compared to that of Ref.~\cite{epjcroma}, while resource usage is approximately similar: more LUTs are used by the CNN while more DSPs were used by our implementation. It is not possible at the present moment to compare directly the resulting trigger efficiency curves since different simulation datasets and detector geometry were used by the two approaches. Another approach~\cite{CMS:2018wav} by the CMS experiment used boosted decision trees as the regression algorithm for measuring muon $\pT$. This approach requires large memory banks to store lookup tables, with a significantly large memory footprint compared to our FPGA implementation of the neural network regression model.

The ATLAS RPC muon trigger system is currently being designed and exact specifications are still being finalised. The overall latency of the future ATLAS hardware trigger system~\cite{CERN-LHCC-2017-020} is fixed at 10~$\mu$s. The latency of the RPC muon trigger logic~\cite{CERN-LHCC-2017-020} is expected to be around 300~ns. It is a small fraction of the total latency because extra time is needed for sending data from the on-detector electronics to the counting room and for data processing by other components of the hardware trigger system. XCVU13P FPGAs are being planned for the RPC trigger system~\cite{CERN-LHCC-2017-020}. This device has about 12,000 DSPs and about 1.8 million LUTs, which is about a factor of four larger resource availability than that of the XCKU060 FPGA. Therefore, our neural network implementation is capable of meeting the latency and resource requirements of the future RPC trigger system.

\section{Conclusions}
\label{sec:sum}

This paper presents a study that aims to improve performance of the future ATLAS hardware muon trigger system in the barrel detector region. A neural network regression model was developed for estimating transverse momentum ($\pT$) and charge of muon trigger candidates. A simplified simulation model of the current ATLAS resistive plate chamber (RPC) detector was developed to generate events for training and testing the regression model. Our model promises to provide more precise measurements of the muon trigger candidate $\pT$, when compared to the current RPC muon trigger. This can improve rejection of low-$\pT$ muon candidates, which are the dominant source of background events accepted by the ATLAS RPC muon trigger.

The latency and resource usage of our FPGA implementation of this neural network are well within the requirements of the future ATLAS hardware muon trigger system. This result opens possibilities for deploying machine learning algorithms for new hardware-based triggers. Verifying these results with a more accurate detector simulation model and developing new FPGA trigger algorithms for exotic particle searches will be the subject of our future work.

\begin{acknowledgements}
This work was supported in part by the National Natural Science Foundation of China, under Grants 11922510 and\\ 11961141014, and by the Fundamental Research Funds for the Central Universities of China, under Grant WK2360000011. We would like to thank Wenqi Lou for the helpful guidance on FPGA implementation using high level synthesis tools.
\end{acknowledgements}

\patchcmd{\bibsetup}{\interlinepenalty=5000}{\interlinepenalty=10000}{}{}
\printbibliography[heading=none]

\end{document}